\documentclass{iaus}
\usepackage{graphicx}

\title[Magellanic Cloud planetary nebulae] 
{The population of Magellanic Cloud planetary nebulae}

\author[Letizia Stanghellini]   
{Letizia Stanghellini$^1$
 }

\affiliation{$^1$National Optical Astronomy Observatory, 950 N. Cherry Ave, Tucson Az 85719}

\pubyear{2008}
\volume{256}  
\pagerange{}
\jname{The Magellanic System: Stars, Gas, and Galaxies}
\editors{Jacco Th. van Loon \& Joana M. Oliveira, eds.}
\begin{document}

\maketitle

\begin{abstract}

In this review we address the progress that has been made toward the understanding the population of 
Magellanic Cloud  planetary nebulae  since the last Magellanic Cloud Symposium. Planetary nebulae in the Clouds are 
not only important as key probes of stellar and ISM evolution in these galaxies, but also reflect the evolution 
of AGB stars and beyond in low-metallicity environments. We present the recent surveys results, including the 
wide fields ground-based search for PNe, the {\it HST} study of the resolves ejecta and their central stars, and the 
{\it Spitzer} analysis of the gas contents of these ejecta. Finally, we show how Magellanic Cloud PNe can be used to
constraint the distance scale of galactic PNe.

\end{abstract}

\firstsection 
\section{Introduction}
Planetary nebulae (PNe) are direct probes of the evolution of low- and intermediate-mass stars (LIMS); LIMS are the
major component of stellar mass in all types of  known galaxies and in the intra-cluster medium, they go through
the asymptotic giant branch (AGB) phase, which is characterized by very high IR luminosities, high mass loss rates, and 
the production of carbon and nitrogen. Through PNe, this important phase of the evolution is made observable across
different galaxy types, and as far
as $\sim$30 Mpc. The importance of AGB stars in galaxies is especially important since they are the major producers of 
nitrogen in the universe, and they supply as much carbon as massive stars. The knowledge of AGB and PN evolution in different
environments, and, in particular, at different metallicities, is essential to soundly constraint the models of stellar and galactic evolution.
 
Planetary nebulae in the Magellanic Clouds have always been of great interests for the metallicity baseline that the Clouds 
offer, Z$\sim0.2-1 ~{\rm Z}_{\odot}$ (\cite[Russell \& Bessell 1989)]{RussellBessell89}, which makes them the benchmark for
AGB studies at low metallicity, essential for the understanding of the integrated light of unresolved galaxies (\cite[Maraston 1995)]{Maraston95}.
While the Magellanic Cloud PNe are typically 50 times farther away than their galactic counterparts,  their distances 
uncertainties are very low, $\sim5\%$ compared to  $\sim 50\%$ or more for galactic PNe (\cite[Stanghellini et al. 2008)]{Stanghellinietal08}, 
making the former the best absolute probes of stellar evolution. Nonetheless, the Magellanic Cloud PNe are close enough to
be studied in much detail, both spectroscopically and via imaging. Furthermore the low selective reddening 
toward the Clouds represent a further advantage for PN studies with respect to the Galaxy. 

\section{Progresses since last IAU Symposium}

In the last decade the field of Magellanic Cloud PNe has advances greatly, especially in several  directions.
The use of systematic
surveys have allowed the discovery and spectroscopic confirmation of many more Magellanic Cloud PNe, more than doubling their number:
About 230 PNe were known in the LMC (\cite[Leisy et al. 1997]{Leisyetal97}) 
at the time of the last Magellanic Cloud IAU Symposium (\cite[Dopita 1999]{Dopita99}), while recently \cite[Reid \& Parker(2006)]{Reidparker06} have 
identify $\sim$700 PN candidates,  $\sim$300 of whose have been spectroscopically confirmed (Reid, this volume). In the SMC, 
\cite[Jacoby \& DeMarco (2002)]{JacobyDeMarco02} analyzed a sampe of $\sim$60 PNe that define a complete sample
6 magnitude down the planetary nebula luminosity function (PNLF) cutoff for the central 2.8 deg$^2$. This makes the SMC the first
galaxy where the features of the PNLF can be studied and interpreter based on a complete sample.

Ground-based spectroscopy is essential not only to confirm a target as a bona fide PN, but also to analyze the plasma and determine the
elemental abundances. In the last decade, thanks especially to the work of \cite[Leisy \& Dennefeld (2006)]{LeisyDennefeld06}, and also 
\cite[Costa et al. (2000)]{Costaetal00}, and \cite[Idiart et al. (2007)]{Idiartetal07},
a large database of PN abundances has been developed for the Clouds. 
But it is especially with space astronomy that the field of Magellanic Cloud PNe have been recently forwarded. 
The extensive use of  the {]it HST}, which has the capability of resolving
them spatially, allowed for the fist time morphological studies of extragalactic PNe, and made their central stars directly observable.
Central stars of Magellanic Cloud PNe have been studied in detail, and from sizable samples, by Villaver et al. (see \cite[Villaver et al. 2006]{Villaver06}),
and, given the known distances of Magellanic Cloud PNe, it had been possible to locate quite accurately these stars on the HST diagram for
direct comparison with the stellar evolutionary tracks. Also from space, UV imaging and spectroscopy allowed the study of carbon
emission lines in Magellanic Cloud PNe (\cite[Stanghellini et al. 2005]{stanghellinietal05}), and the derivation of carbon abundances
in LMC PNe, a clear signature of the evolution of LIMS. 
As {\it Spitzer} became available, the IR spectra of Magellanic Cloud PNe became observable, affording dust studies 
both in imaging (\cite[Hora et al. 2008]{horaetal08}) and spectroscopy (\cite[Stanghellini et al. 2007]{Stanghellinietal07}); furthermore, {\it Spitzer} 
spectroscopy allowed more insignt on elemental abundance analysis by including the IR transition into the calculations (\cite[Bernard-Salas et al. 2004]{Bernardsalas04}). 
Another important aspect that has advanced in the field of magellanic Cloud PNe is the publication of new sets of stellar models,
both synthetic (\cite[Marigo 2001]{Marigo01}) and evolutionary (e.g., \cite[Karakas \& Lattanzio (2008)]{KarakasLattanzio08}), based on
initial conditions that reflect those of the metallicities in the Magellanic Cloud populations. These models give the yields of the element of stellar
evolution in relation to the few final thermal pulses, when the PN is ejected, thus are readily comparable with the PN observations.

The wealth of new data and models make the Magellanic Cloud PNe the ideal laboratory to study stellar populations and evolution at
various metallicities; these have been used to explore, and trying to answer, some open question in the field on PNe, in particular, the focus has been on 
(1) 
the different PN morphologies, and how these originate and evolve; (2) the chemistry of PNe as they evolve in different metallicity environments, and
how do they contribute to cosmic recycling; (3) the evolution of the PN central stars (CS); (4) the PNLF in the Magellanic Clouds, and how can it
be used to constraint and scale the extragalactic distance scale; (5) the role of dust in PNe as agent in PN ejection, evolution, and morphology, and the 
nature of dust in PNe of different metallicities; and finally (6) the use of Magellanic Cloud PNe as calibrator of the Galactic PN distance scale. In the
next few sections we will explore several of these topics.

\section{Magellanic Cloud PN morphology, and the evolutionary connection}

Planetary nebula morphology displays the evolutionary history of the final phases of the LIMS life, and it has shown to be correlated to
both the evolution population of the LIMS progenitor. Magellanic Cloud PNe are only 0.5 arcsec across on average, thus spatially resolvable only 
with observations from space. In the last decade many samples of Magellanic Cloud PN images have been acquired with the {\it HST},
forming a large database to study the morphological evolutionary connection (\cite[Shaw et al. 2001]{shawetal01}, \cite[Stanghellini et al. 2002]{Stanghellinietal02},
 \cite[Shaw et al. 2006]{Shawetal06}). These studies include 114 LMC and 35 SMC PNe, representing approximately 2/3 of all bright Magellanic Cloud PNe, 
 and populate  the 5 bright magnitude bins of the PNLF (in terms of m$_{\lambda5007}$). Morphology of Magellanic Cloud PNe can be divided into major classes,
round, elliptical, bipolar, and bipolar core or ring PNe , just as their Galactic counterparts. In Table 1 we show the statistics of the morphological types in 
Magellanic Cloud PNe, where we enlarge the sample above with the PNe already observed with the {\it HST} by Dopita, Vassiliadis, and 
collaborators (see \cite[Dopita (1999)]~. We note that asymmetric PNe (as we call the bipolar PNe, and also the ones showing asymmetries such as bipolar cores)
 are much more common in the LMC than the SMC, and, in particular, the number of bipolar PNe in the SMC is very low.

\begin{table}
  \begin{center}
  \caption{Magellanic Cloud PN morphology}
  \label{tab1}
 {\scriptsize
  \begin{tabular}{lrr}\hline 
  & LMC& SMC\\
  &&\\
  round& 29$\%$& 35$\%$\\
  elliptical& 17$\%$& 29$\%$\\
  bipolar& 34$\%$& 6$\%$\\
  bipolar core& 17$\%$& 24$\%$\\
  point-symmetric& 3$\%$& 6$\%$\\
   \end{tabular}
  }
 \end{center}
   \end{table}

The origin of PN morphology can be ascribed to the mechanism of mass ejection at the tip of the AGB and to the condition of the circumstellar
medium at that evolutionary phase: The round and most elliptical shapes can be formed, according to hydrodynamic models, via ballistic expansion, while to create
the bipolar shape a equatorial gas and dust enhancement needs to be present at the time of the envelope ejection . What creates the conditions for the
enhancement is still controversial in PN studies, we will see late in this review how the dust and gas studies of Magellanic Cloud PNe have 
greatly helped to answer this hot issue of PN evolution. Most of the models for highly asymmetric PN evolution involve either rotation and magnetic fields
(\cite[Garcia-Segura 1997]{garciasegura97}), or common-envelope (CE) processes (Morris 1981; Soker 1998). The former set of models agree with a 
progenitor mass-dependency of the morphological evolution, while the latter set of models, those involving the CE evolution, are mass-independent, since
the chance of close binary evolution probably doe snot depend of the mass of the progenitor. 

The Magellanic Cloud data have helped to clarify that
the process forming bipolar PNe is necessarily mass-dependedn, in most cases. In fact, if bipolarity is related to more massive LIMS,
these PNe should carry the signature of more massive progenitor's evolution in their elemental
abundance. From stellar evolution we know that the third dredge-up, and the hot bottom burning process, only occur in the most massive LIMC 
(M/M$_{\odot}>$3 to 4). These processes have the net effect to reduce carbon and enhance nitrogen (and N/O) in progenitors whose
metallicities are those of the LMC and the SMC as well as at solar metallicity (Marigo 2001; Karakas \& Lattanzio 2008). 
In Table 2 the average abundances of the key evolutionary elements for those Magellanic Cloud PNe whose morphology 
have been classified via the {\it HST} images are shown. The sources for the abundances are from \cite[Leisy \& Dennefeld (2006)]~, excluding uncertain values. carbon
abundances are also from \cite[Stanghellini et al. (2005)] ~ for the LMC PNe, and from new ACS prism spectra (Stanghellini et al., in preparation).
The averages are given for the whole sample found in the literature, and also for the morphological groups of round and elliptical, and bipolar core and bipolar, PNe. 
For the LMC, where the sample is large enough, the table lists  separately the bipolar PNe, those with well-defined bipolar lobes.

It is very evident that the PN chemistry closely follows their morphology. A look at the LMC averages in Table 2 shows carbon is depleted
in the asymmetric PNe, whereas nitrogen is strongly enriched. A similar trend is observed in the SMC (note that in the SMC there aren't bipolar PNe whose
carbon and nitrogen have been accurately measured). These trends are found in all galaxies where morphological studies are possible, independent on metallicity. 
Figure 1 shows the averages and ranges of the N/H ratios in SMC (Z=0.004), LMC (Z=0.008), and Galactic (Z=0.016) PNe, plotted against the galaxy metallicity. 
The Galactic data are from \cite[Stanghellini et al. (2006)]{Stanghellinietal06}. There is no doubt that, whatever makes the asymmetric PNe acquire their shape, it has
to be closely correlated with nitrogen yield, and this, in turn, is correlated with the progenitor mass, independent on metallicity. The mass-dependent mechanism to form bipolar PNe should work in most cases, although there are a few bipolar Pne with low nitrogen (and/or high carbon) abundance. In these cases, the mechanism forming the 
bipolar is mass-independent, and could be the CE evolution after the close binary interaction. Interestingly, the first evolutionary models show that yields of 
close binary evolution do not enhance nitrogen nor deplete carbon (\cite[Izzard 2007]{Izzard07}), thus can not describe the formation of the 
majority of SMC, LMC, and Galactic asymmetric PNe.

\begin{table}
  \begin{center}
  \caption{Evolutionary connection}
  \label{tab2}
 {\scriptsize
  \begin{tabular}{lrr}\hline 
  & LMC& SMC\\
  &&\\
$<$C/H$>\times10^4$& & \\
whole sample&	2.49$\pm$2.18&  3.71$\pm$3.66\\
round, elliptical&	3.96$\pm$2.00&	4.55$\pm$3.86\\
bipolar core, bipolar&	2.10$\pm$2.13&   2.00\\
bipolar&   0.47$\pm$0.45& $\dots$ \\
&&\\
$<$N/H$>\times10^4$& & \\
whole sample&  1.48$\pm$1.75& 0.29$\pm$0.33\\
round, elliptical&	1.00$\pm$1.35& 0.18$\pm$0.68\\
bipolar core, bipolar& 2.30$\pm$1.99& 0.65$\pm$0.87\\	
bipolar&	2.68$\pm$2.11& $\dots$ \\	
&&\\
$<$N/O$>$&&\\
round, elliptical&  0.57$\pm$0.89& 0.12$\pm$0.09\\
bipolar core, bipolar& 1.31$\pm$1.45& $\dots$ \\
bipolar& 1.54$\pm$1.61& $\dots$ \\

   \end{tabular}
  }
 \end{center}
  \end{table}

\begin{figure}[b]
\begin{center}
 \includegraphics[width=3.4in]{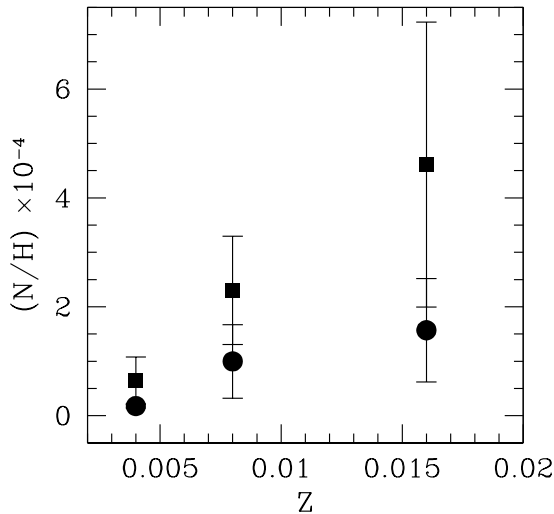} 
 \caption{Average nitrogen abundances in the populations of SMC, LMC, and Galactic PNe with known morphology, plotted
 against the mean metallicity. Averages
 are plotted with filled circles and  squares for symmetric (round and elliptical) and asymmetric (bipolar core and bipolar) PNe
 respectively. Bars represent data ranges. }
   \label{fig1}
\end{center}
\end{figure}

\section{Abundances in Magellanic Cloud PNe, and metallicity gradients}

The abundance of oxygen, neon, and other $\alpha$-elements in the Magellanic Cloud PNe allow the study of the 
elements at the time of the progenitor formation, since oxygen and neon are mostly produced in massive stars in primary
nucleosynthesis, and not in LIMS.
In Figure 2 we show the distribution of oxygen and neon in the LMC (open symbols) and the SMC (filled symbols) PNe, where 
morphological types have been also coded. The relation is tight, as expected from elements in lockstep evolution. 
The average Ne/O ratio is $<$Ne/O$>$=0.17$\pm$0.09 both in the LMC and the SMC, while it is 0.27 in
Galactic PNe (Stanghellini et al. 2006). The lower ratio at low metallicity shows that oxygen and neon abundances do not always scale
with metallicity in lockstep. 
The abundance distribution of $\alpha$elements in the Clouds, derived from the PN population, has not shown a clear metallicity gradient,
nor a relation between metallicity and the location of the star forming regions in neither the LMC nor the SMC. The morphological type distribution across
the LMC has also been studied (Stanghellini 2002) to show no particular morphological segregation, in agreement with a short crossing time of LMC
compared to the timeframe of LIMS evolution.

\begin{figure}[b]
\begin{center}
 \includegraphics[width=3.4in]{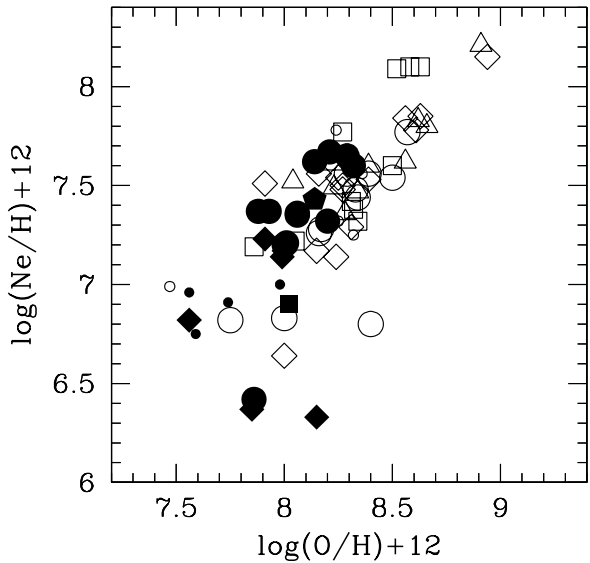} 
 \caption{Neon vs. oxygen abundances, in log scale, for LMC (open symbols) and SMC (filled symbols) PNe. Shape of symbols 
 indicate morphological types: Circles=round, diamonds=elliptical, triangles=bipolar core, squares=bipolar, and small circles=unknown morphology.}
   \label{fig2}
\end{center}
\end{figure}

\section{Central stars of Magellanic Cloud PNe}
The direct images of PN central stars (CS) is only possible through space imaging. Villaver et al. (2003, 2004, 2007) has analyzed a series of databases of LMC and SMC
CS images acquired with STIS and WFC2 on the {\it HST}, and, by measuring their luminosity and temperature (with the aid of ground-based spectroscopy as well),
estimate their masses. Villaver et al. (2007) determined that $<$M/M$_{\odot}>_{\rm LMC}$=0.65$\pm$0.07 and very similar for the SMC, 
which is slightly higher than what is generally estimated in Galactic 
PNe. The possible explanation here is that, at lower metallicity, the mass loss at the TP-AGB is less efficient, thus the remnant stellar mass are higher in the Clouds than in the
Galaxy (Villaver et al. 2003, 2004).  Interestingly, Villaver et al. found no clear cut relation between stellar mass and PN morphology in the magellanic Cloud sample, 
as opposed to Galactic PNe (Stanghellini et al. 1993, 2003). It seems like in the Cloud PNe, by the time the superwind is over there is no longer
memory of the initial mass. 

The CS of the Magellanic Cloud PNe analyzed by Villaver are among the few PNe with known distances. As such, they offer the opportunity to test the PNLF, and, in particular, to
see which PNe populate the high luminosity cutoff of the PNLF, which is used to calibrate the extragatactic distance scale. In Figure 3 we show the relation between
the measured radii and the CS luminosities of the CS in the Clouds, where the symbols keys are the same as in Figure 2. We see that the brightest CS are those 
hosted by compact PNe in both the LMC and the SMC, with R$_{\rm phot}<$0.5 pc. This is expected, as the evolution of central stars (on the HR diagram) follows a luminosity plateau
right after PN ejection, and then evolves toward the WD cooling line. Nonetheless, this is shown with data for the first time in the Clouds, since the 
distance sof Galactic PNe are too uncertain for these types of comparisons.  Central stars have mass between $\sim$0.55 and 1.4 M$_{\odot}$, and the 
stars populating the bright end of the PNLF are the intermediate mass stars in the population of the Magellanic Clouds.

\begin{figure}[b]
\begin{center}
 \includegraphics[width=3.4in]{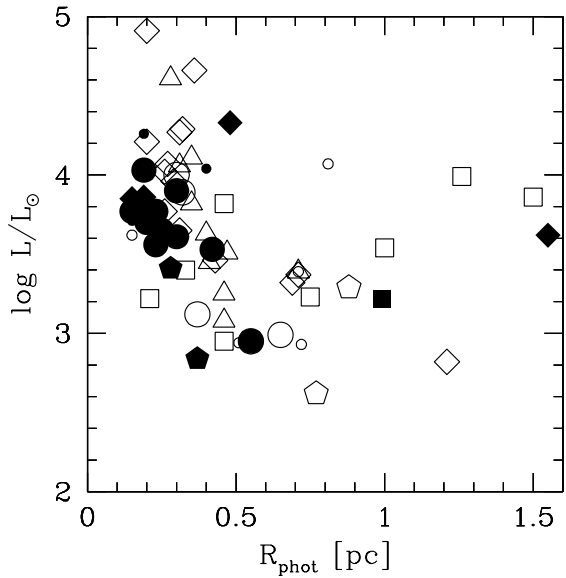} 
 \caption{The measured CS luminosities vs. the PN physical radii, for LMC (open symbols) and SMC (filled symbols) PNe. Shape of symbols 
 indicate morphological types as i Fig.~2.}
   \label{fig3}
\end{center}
\end{figure}

\section{Dust in Magellanic Cloud PNe}

The study of dust in the Magellanic Cloud PNe through mid-IR spectroscopy  had been possible only with the availability of the {\it Spitzer} space telescope, as only the few brightest PNe
were within reach of the earlier technology. The IRS/{\it Spitzer} spectra are also extremely important for the abundance analysis ([cite{Bernard-Salas et al 2004]{Bernardsalas04}), but here we will focus on the dust. 
The importance of studying dust in PNe, especially in the Magellanic Clouds, is that mass-loss at the AGB tip is still not completely understood, but theoretical
work indicates that dust and metallicity are essential keys in the mass-loss efficiency (Willson 2000). While mass-loss may occur in the absence of dust,
it is believed that in most cases the pressure on the dust grains produces the strong mass-loss at the AGB tip (Willson 2000), and that mass-loss efficiency in this
case is directly proportional to metallicity, through the dependence of the absorption coefficient.

The existing observations seem to agree with this correlation: there are fewer obscured AGB in the Magellanic Clouds than in the Galaxy (Groenewegen 2000). Also,
the C-rich to O-rich ratio of AGB stars is higher at lower metallicity (Cioni \& Habing 2003). Finally, as seen above, there are fewer aspheric PNe in the SMC than in the LMC. On these basis, Stanghellini et al. (2007) examine a homogeneous set of IRS spectra to determine the IR/dust properties of the Magellanic Cloud PNe whose
morpohlogy had been previously determined with the {\it HST} images. The analyzed spectra display three types of behavior: half of them are featureless, except for the nebular emisison line and a weak dust continuum, while the other half show dust features in the form of solid state emission features. In most cases these features 
are recognized as carbon-rich dust features such as SiC and PAH, while in only three PNe the trace of oxygen-rich dust was disclosed. 

In Figure 4 we show how the dust features in the IRS spectra correlate with the nebular gas. Both panels show the IR luminosity, derived from a black-body fit of
the dust continua in the feature- and line-subtracted spectra, versus the (gas) carbon abundance of the PNe. The left panel characterizes the dust properties
of the PNe, while the right panel indicates the PN morphology.  From the left panel we infer that all featureless spectra (triangles) are in the lower luminosity part
of the diagram. It is also evident that the carbon-rich dust PNe (diamonds) correspond to high carbon abundances, and vice versa the oxygen-rich dust PNe(squares) 
are those with low carbon. There is then a direct relation between carbon in the dust and the gas of Magellanic Cloud PNe. Note that we did not distinguish LMC from SMC PNe in this plot, to avoid confusion. The right panel shows that all round and elliptical PNe (circles and diamonds) correspond to carbon-rich dust, or featureless, 
IRS spectra. There is a good correspondence between round/elliptical (symmetric) and carbon-rich dust PNe, and 
none of these are bipolar, and none of the oxygen-rich dust PNe  are round or elliptical.

There is also a strong difference between dust features in the Magellanic Cloud PN populaiton and the galactic one, in that all observed galactic PN show solid state features (Garcia-Lario et al., in preparation). The ratio of carbon-rich dust over oxygen-rich dust PNe is $\sim$11 in the SMC, $\sim$4.5 in the LMC, while it is 
estimated to be close to unity in the Galaxy (Garc'a-Lario et al), indicating that the population metalliticy has enormous impact on AGB dust formation.

\begin{figure}[b]
\begin{center}
 \includegraphics[width=6.6in]{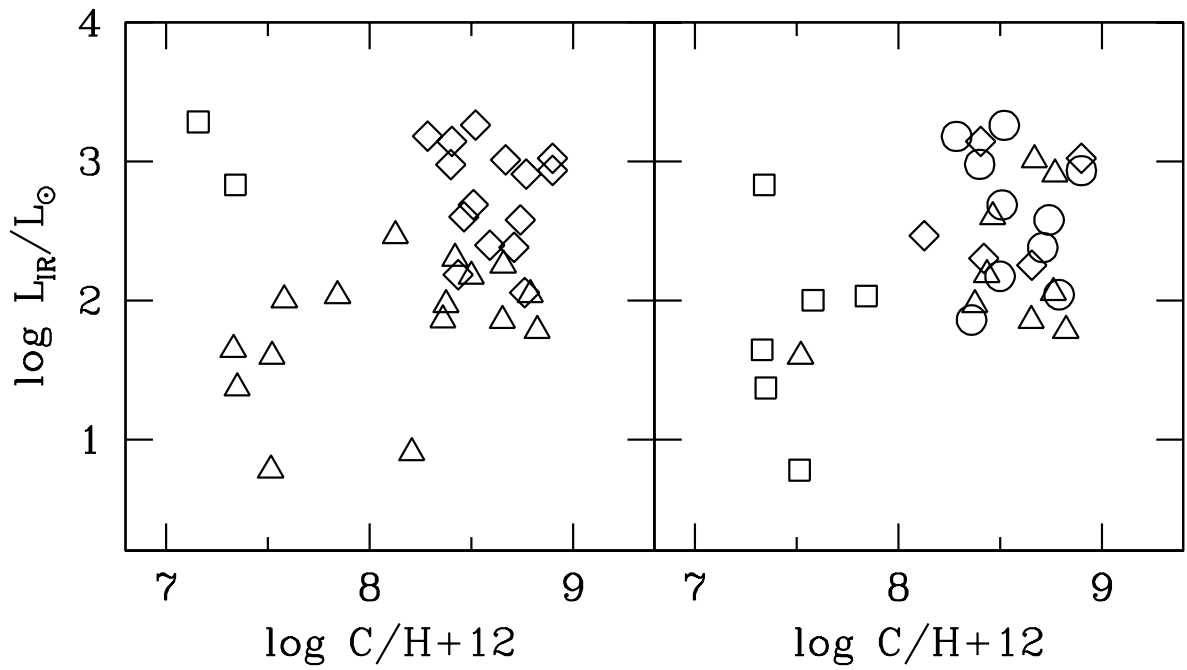} 
 \caption{IR Luminosity vs. carbon abundance of Magellanic Cloud PNe. Left panel: dust properties (triangles: no dust features; diamonds: 
 carbon-rich dust; and squares: oxygen-rich dust PNe). Righ panel: morpohlogy (circles: round; diamonds: elliptical; squares: biplar, triangles: bipolar core
 PNe).}
   \label{fig4}
\end{center}
\end{figure}

\section{Magellanic Cloud PNe as distance calibrators}

There are $\sim$1800 PNe in the galaxy , but for only a a few accurate distances have been determined. 
Statistical distances based on physical correlations of nebular parameters  can be calibrated  with the newly observed Magellanic Cloud PNe, a 
sample of more than 100 calibrators with very well known distances.

\section{Future endeavors}


\begin{thebibliography}{}

\bibitem[Russell \& Bessell (1989)]{RussellBessell89} {Russell, S.~C., \& Bessell, M.~S.\ 1989, ApJS, 70, 865 }
\bibitem[Maraston (1995)]{Maraston95}{Maraston, C.\ 2005, MNRAS, 362, 799 }
\bibitem[Stanghellini et al. (2008)]{Stanghellinietal08}{Stanghellini, L., Shaw, R. A., \& Villaver, E., 2008, ApJ, in press}
\bibitem[Leisy et al. 1997]{Leisyetal97}{Leisy, P., Dennefeld, M., Alard, C., \& Guibert, J.\ 1997, A\&AS, 121, 407}
\bibitem[Dopita 1999]{Dopita99}{Dopita, M.~A.\ 1999, New Views of the Magellanic Clouds, 190, 332}
\bibitem[Reid \& Parker(2006)]{Reidparker06}{Reid, W.~A., \& Parker, Q.~A.\ 2006, MNRAS, 373, 521}
\bibitem[Jacoby \& DeMarco (2002)]{JacobyDeMarco02}{Jacoby, G.~H., \& De Marco, O.\ 2002, AJ, 123, 269}
\bibitem[Leisy \& Dennefeld (2006)]{LeisyDennefeld06}{Leisy, P., \& Dennefeld, M.\ 2006, A\&A, 456, 451}
\bibitem[Costa et al. (2000)]{Costaetal00}{Costa, R.~D.~D., de Freitas Pacheco, J.~A., \& Idiart, T.~P.\ 2000, A\&AS, 145, 467}
\bibitem[Idiart et al. (2007)]{Idiartetal07}
\bibitem[Villaver et al. 2006]{villaver06}
\bibitem[Stanghellini et al. (2005)]{stanghellinietal05}
\bibitem[Hora et al. (2008)]{horaetal08}
\bibitem[Stanghellini et al. (2007)]{Stanghellinietal07}
\bibitem[Bernard-Salas et al. (2008)]{Bernardsalas08}
\bibitem[Bernard-Salas et al. (2004)]{Bernardsalas04}




\end{thebibliography}
\end{document}